\documentclass[letterpaper,twocolumn]{jpsj3}
\usepackage{txfonts}
\usepackage[dvipdfmx]{graphicx}
\usepackage{color}
\usepackage{ulem}

\title{Novel Easy-Axis Switching through Metamagnetism in CeSb$_2$}

\author{
Atsushi Miyake$^1$\thanks{atsushi.miyake.b6@tohoku.ac.jp}, 
Ryuta Hayasaka$^1$,
Hiroto Fukuda$^1$,
Masaki Kondo$^2$,
Yuto Kinoshita$^2$,
Dexin Li$^1$, 
Ai Nakamura$^1$, 
Yusei Shimizu$^1$, 
Yoshiya Homma$^1$, 
Fuminori Honda$^{3}$, 
Masashi Tokunaga$^2$, and
Dai Aoki$^{1}$
}

\inst{
$^1$Institute for Materials Research, Tohoku University, Oarai, Ibaraki 311-1313, Japan \\
$^2$Institute for Solid State Physics (ISSP), The University of Tokyo, Kashiwa, Chiba 277-8581, Japan\\
$^3$Central Institute of Radioisotope Science and Safety, Kyushu University, Fukuoka, 819-0395, Japan\\ 
} 

\abst{
A novel magnetic field-induced switching of the magnetization easy axis has been discovered in the layered compound CeSb$_2$, which crystallizes in an orthorhombic structure with nearly identical lattice constants along the a- and b-axes, giving it a tetragonal-like appearance.
When a magnetic field is applied along an orthorhombic in-plane axis at 4.2~K, magnetization increases abruptly around 34~T, followed by a hysteresis loop upon decreasing the field. 
Subsequent measurements reveal a significantly enhanced magnetization, indicative of a switch of the magnetization easy axis. 
Conversely, the other orthorhombic in-plane axis becomes the magnetization hard axis.
Surprisingly, the easy axis is switchable by changing the field direction to the other orthorhombic in-plane axis.
Moreover, this memory effect is stable up to room temperature.
Polarized light microscope images have visually revealed domain formation of the as-cast samples and domain rearrangement by magnetic fields.
This intriguing axis-conversion phenomenon is a novel magnetic shape memory effect for heat cycles up to room temperature. 
It is attributed to the specific in-plane Ce-{\it pantograph} networks in CeSb$_2$.
}
\begin{document}
\maketitle
In rare earth and/or uranium intermetallics, one of the intriguing phenomena induced by magnetic fields is a metamagnetic transition (MMT), which is the nonlinear increase in magnetization as a function of the magnetic field.
Several types of MMTs have been well investigated in the $f$-electron systems.
One well-known example is a pseudo-MMT (metamagnetic-like crossover) observed in Ce-based paramagnetic (PM) metals, such as CeRu$_2$Si$_2$ \cite{Haen_1987} and CeNi$_2$Ge$_2$ \cite{Fukuhara_1998, Miyake_2017}.
First-order MMT in PM systems is also known, for example, in UTe$_2$ \cite{Ran_2019, Miyake_2019, Miyake_2021} and itinerant metamagnets \cite{Goto_2001}.
The MMTs, the field-induced polarized PM phase from the magnetically ordered phase, are also of the first order \cite{Aoki_2013}.
Although these first-order transitions show hysteresis behavior against the fields, the systems recover to the original zero-field state by removing the fields.
For CeSb$_2$, sequential first-order MMTs and the complex magnetic phase diagram have been intensively studied \cite{Budko_1998, Kagayama_2000, Zhang_2017, Liu_2020, Trainer_2021}. 
The recent discovery of pressure-induced unconventional superconductivity observed in a high-pressure phase of CeSb$_2$ renewed interests \cite{Squire_2023}.

Another type of MMT is magnetization easy-axis conversion, which was first observed in an orthorhombic system DyCu$_2$ \cite{Hashimoto_1990, Hashimoto_1994,Sugiyama_1997, Yoshida_1998, Loewenhaupt_1998}.
This MMT is irreversible in the magnetic field scan.
When the magnetic fields are applied along the orthorhombic magnetization-hard $c$ axis, magnetization increases suddenly to the easy $a$-axis value around 13~T.
The magnetization for the demagnetization process traces the $a$-axis magnetization.
This converted state in DyCu$_2$ is retained unless the sample is heated above 100~K or a magnetic field is applied along the original $a$ axis \cite{Hashimoto_1994}.
Following this discovery, other isomorphs $R$Cu$_2$ ($R$ = rare earth ions) were intensively studied \cite{Sugiyama_1995, Settai_1995, Ahmet_1996,Takeuchi_1996, Sugiyama_1998, Settai_1998, Abliz_1997, Andreev_1998, Sugiyama_1999, Svoboda_1999, Loewenhaupt_2000, Sugiyama_2003, Raasch_2006}. 
The flip of the quadrupole moment is discussed as an origin of the conversion phenomena in $R$Cu$_2$ \cite{Settai_1998, Yoshida_1998}.
In this Letter, we will introduce a new member, CeSb$_2$, showing the magnetization easy-axis switching by magnetic fields.

\begin{figure}
\begin{center}
\includegraphics[width=1\hsize]{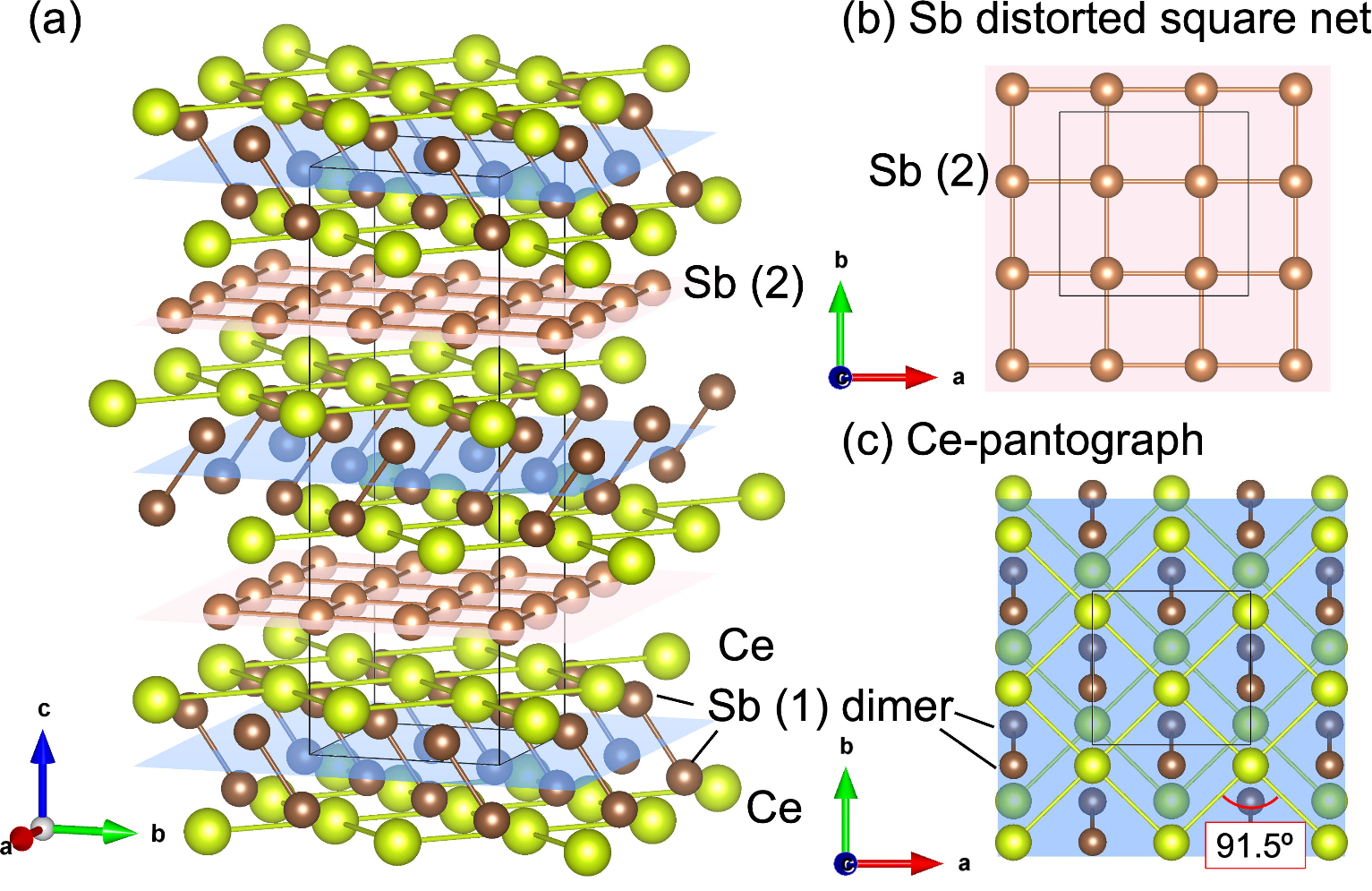}
\end{center}
\vspace{-0.5cm}
\caption{(Color online)
Crystal structure of CeSb$_2$ \cite{Wang_1967} illustrated using $VESTA$ \cite{Momma_2011}.
The thin lines represent a unit cell.
An overview of the structure is given in (a).
(b) quasi-square Sb(2) lattice. 
[The light-red shaded $c$ plane ($z$ = 1/4 and 3/4) in (a)] 
(c)The nearest neighbor Ce atoms form the pantograph networks, which sandwich the Sb dimer [The light-blue shaded $c$ plane ($z$ = 0 and 1/2) in (a)].
}
\vspace{-1.0cm}
\label{structure}
\end{figure}

CeSb$_2$ crystalizes in the layered orthorhombic SmSb$_2$-type crystal structure (No. 64, $Cmce$)\cite{Wang_1967} (Fig.~\ref{structure}).
The lattice constants of $a$ and $b$ axis are slightly different by $\sim$2.8\%; $a$ = 6.295 $\AA$ and $b$ = 6.124 $\AA$.
The Sb and the Ce$\text{--}$Sb block layers stack along the $c$ axis.
The characteristic arrangements appear in each layer, owing to the nearly tetragonal symmetry.
The Sb atoms form a rectangular lattice, slightly distorted from the square lattice [Fig.~\ref{structure}(b)].
The nearest neighbor Ce atoms form rhombuses, networking within the $ab$ plane.
We named this $ab$-plane Ce arrangement the Ce-{\it pantograph} networks, which sandwich the Sb$_2$ dimer.
This unique structure plays a crucial role in the axis conversion.
Reflecting the two-dimensional layered structure of CeSb$_2$, magnetism is highly anisotropic.
The magnetic moments align within the $ab$ plane, and the $c$ axis is the magnetization hardest axis \cite{Budko_1998}. 
For the applied magnetic field $H$ along $a$ and $b$ axis, CeSb$_2$ has complex magnetic phase diagram below $T_{\rm N}$ = 15.5~K and 2~T \cite{Budko_1998, Zhang_2017, Liu_2020, Trainer_2021}.
Multiple MMTs occur at low temperatures, followed by the magnetization saturation behavior.

We found an easy-axis-conversion phenomenon in CeSb$_2$ by applying magnetic fields along the in-plane principal axes far beyond the critical field of antiferromagnetism.
The field direction becomes a magnetization easy axis.
Surprisingly, the easy axis is switchable by changing the field direction to the other orthorhombic in-plane axis.
In contrast to $R$Cu$_2$, the axis conversion is reserved at least up to room temperature, the highest temperature studied here.
We also visualized twinning of the as-cast sample using a polarizing light microscope.
The axis conversion accompanies domain rearrangement, i.e., detwinning.
This phenomenon is regarded as a magnetic shape memory effect.
We discuss possible scenarios for the easy-axis switching observed in CeSb$_2$.

\begin{figure}[t]
\begin{center}
\includegraphics[width=1\hsize]{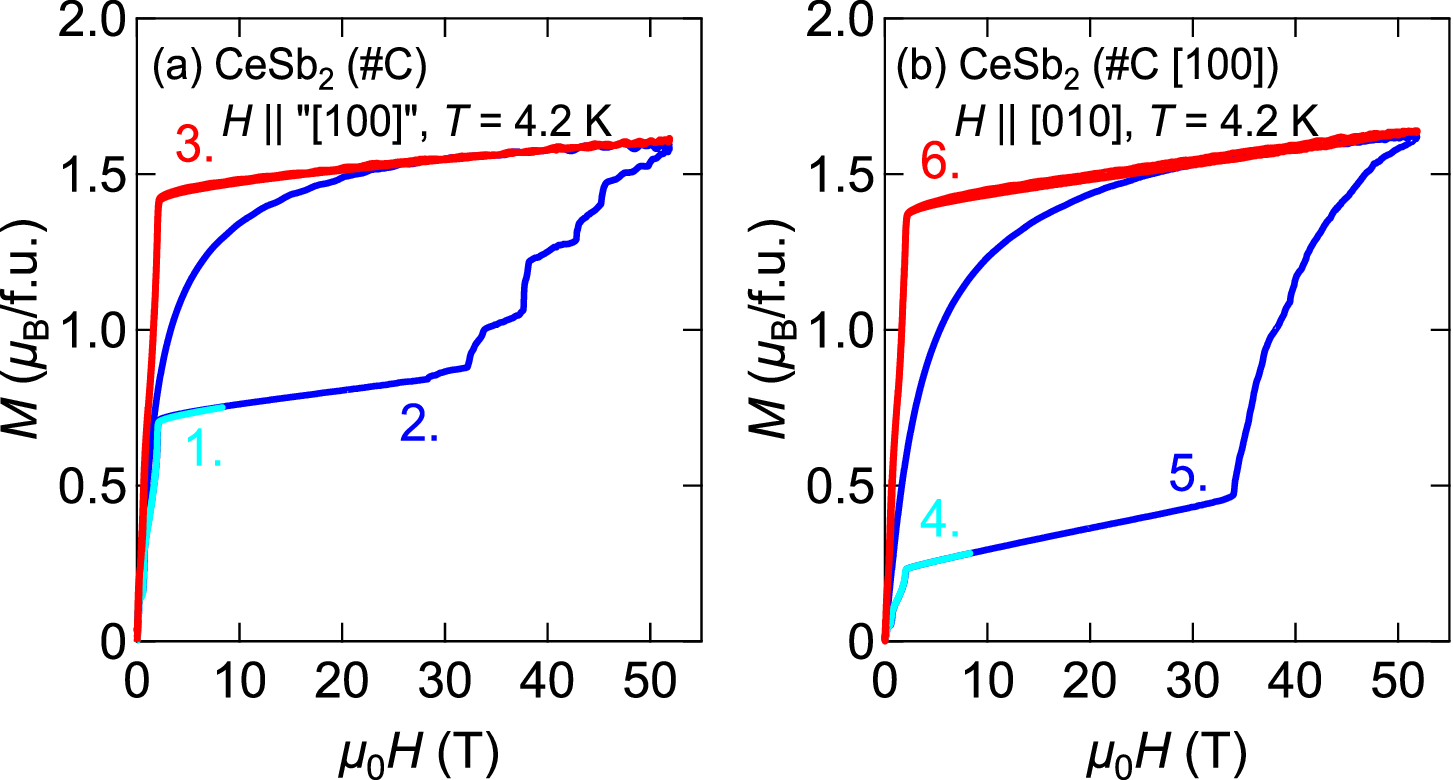}
\includegraphics[width=1\hsize]{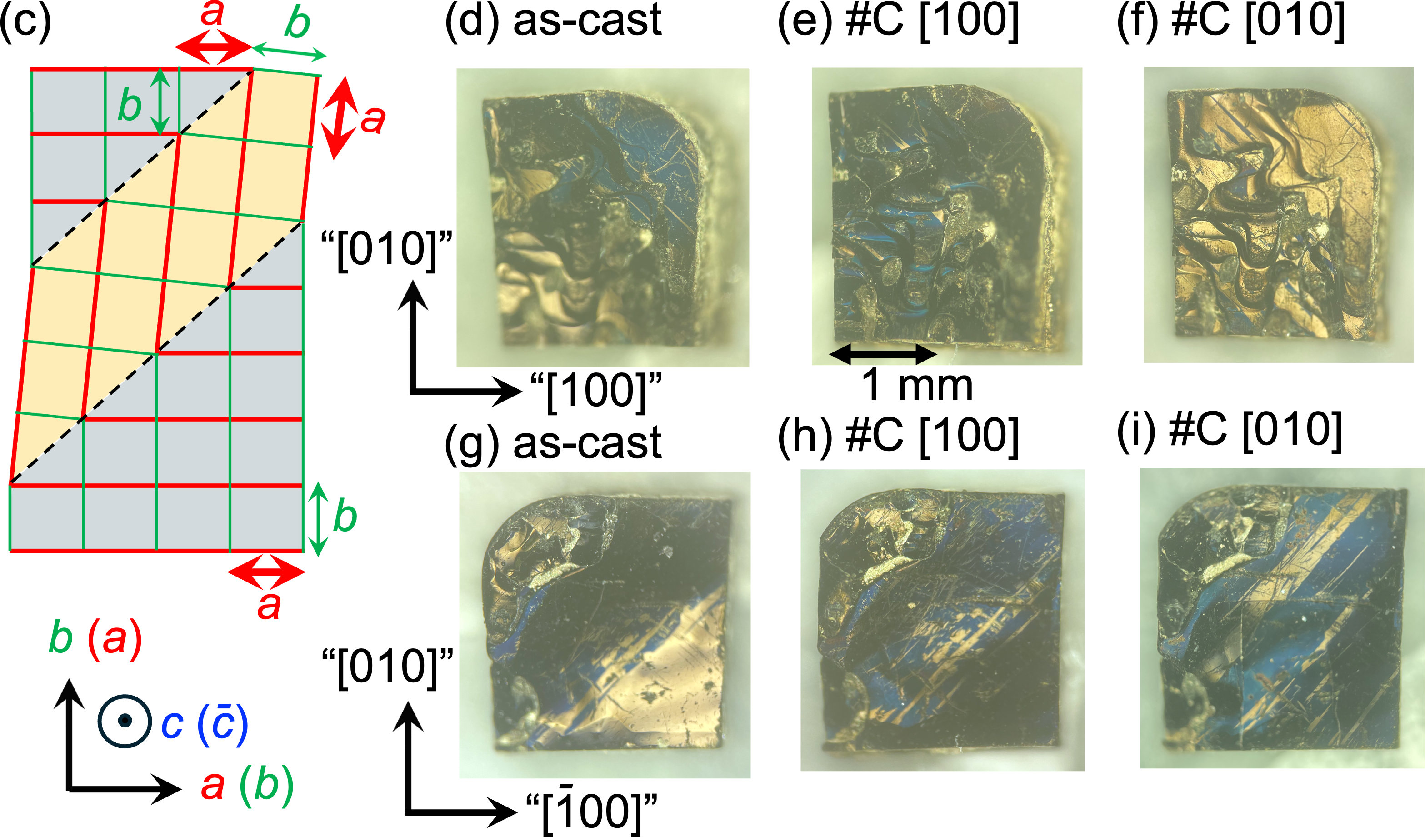}
\end{center}
\vspace{-0.5cm}
\caption{(Color online)
(a) $M(H)$ curves of the as-cast CeSb$_2$ (\#C) at $T=4.2$~K for $H~||$~``[100]''.
The numbers indicate the sequence (shot).
(b) $M(H)$ curves of the sample \#C [100] for $H~||~[010]$ at 4.2~K.
(c) Schematic drawing of the possible domain configuration. 
The dashed lines are the domain walls: diagonals of $ac$ and $bc$ planes.
The polarizing microscope images of the $c$ plane: (d) as-cast, (e) after shot 3, and (f) after shot 6. 
Similarly, (g), (h), and (i) are images of opposite sides of (d)-(f).
The definition of the field directions is also shown.
}
\vspace{-1cm}
\label{H_domain}
\end{figure}

Single crystals of CeSb$_2$ were grown by the Sb-flux method \cite{Budko_1998, Borsese_1981, Zhang_2021}.
The obtained large single crystals were oriented by the Laue photographs and were cut using a spark cutter.
It is noted that the as-cast samples are twinned, and thus, the orthorhombic [100] and [010] axes cannot be defined macroscopically [see Fig.~\ref{H_domain}]. 
The axis determined is distinguished as ``[100]'' with double quotation marks.
The resistivity measurements revealed the very high-quality single crystals with the residual resistivity ratio $\sim$400.
Magnetization measurements were performed using a SQUID magnetometer (MPMS) up to 5~T and down to 2~K\cite{suppl}.
The higher fields of up to approximately 50~T were applied using a pulse magnet with typical durations of $\sim$36 ms, installed at the International MegaGauss Science Laboratory at the ISSP of the University of Tokyo.
High-field magnetizations were measured by a conventional method using coaxial pick-up coils.  
The samples' domains were visualized using a polarizing light microscope at ambient conditions.

\begin{figure}[h]
\begin{center}
\includegraphics[width=0.9\hsize]{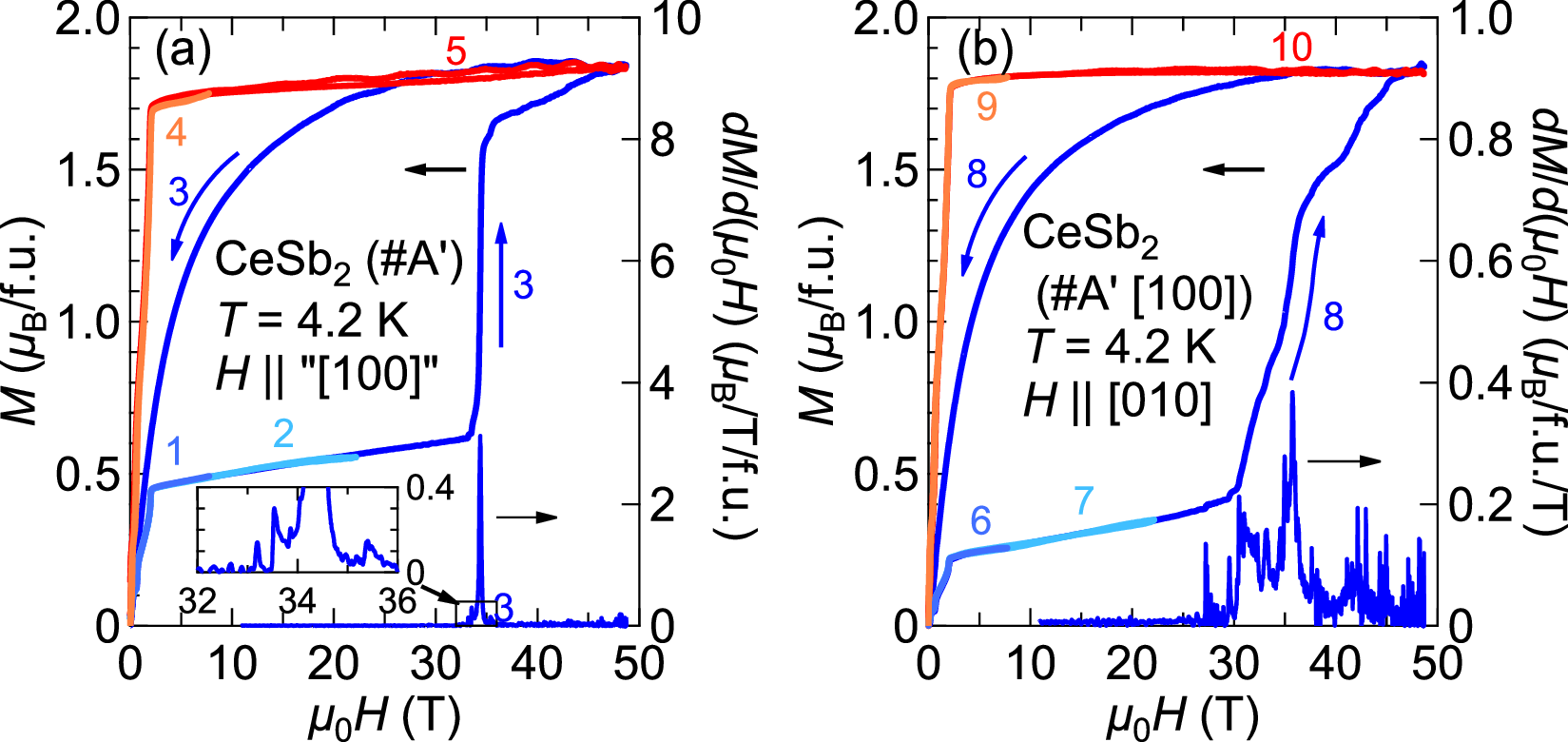}
\end{center}
\begin{center}
\includegraphics[width=0.9\hsize]{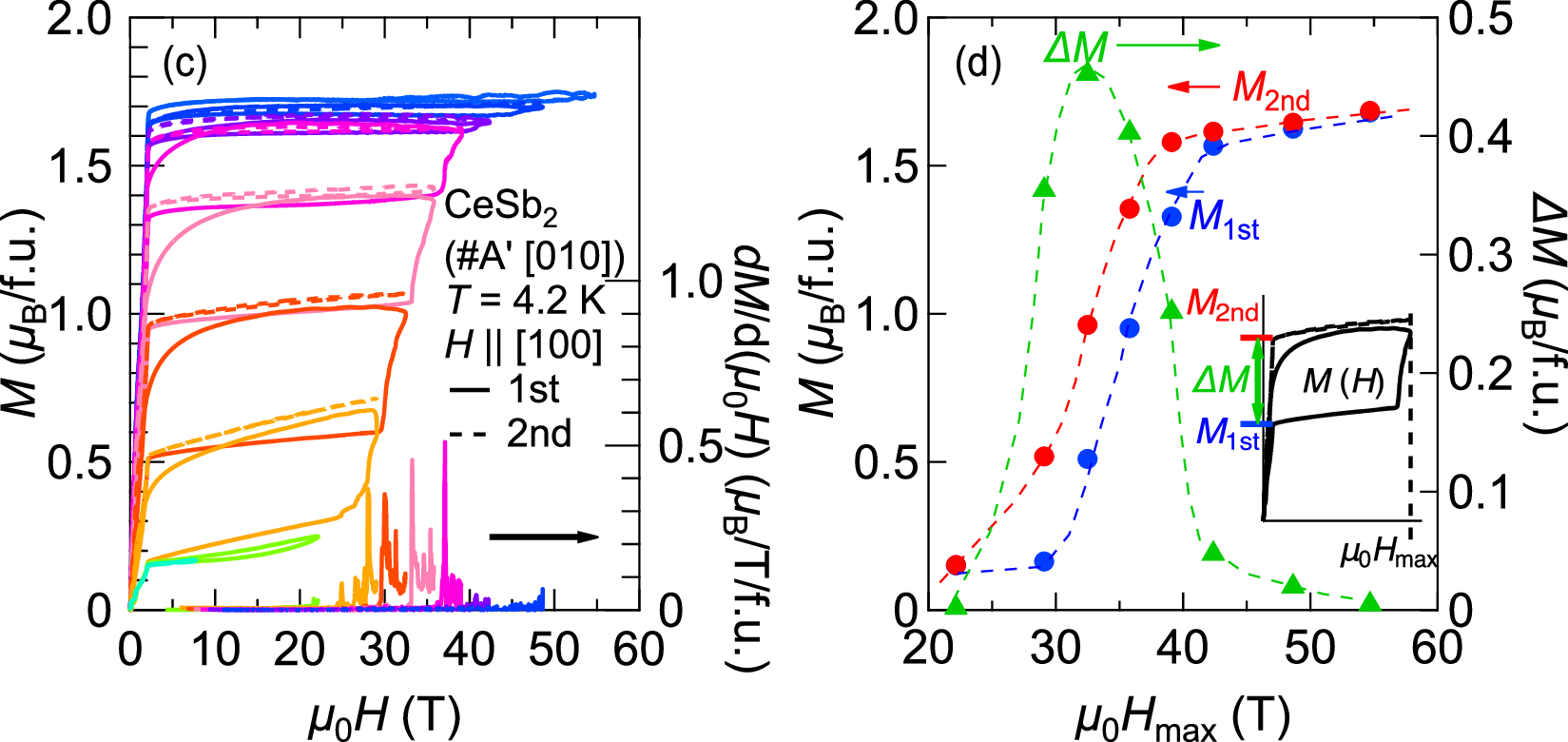}
\end{center}
\begin{center}
\includegraphics[width=0.9\hsize]{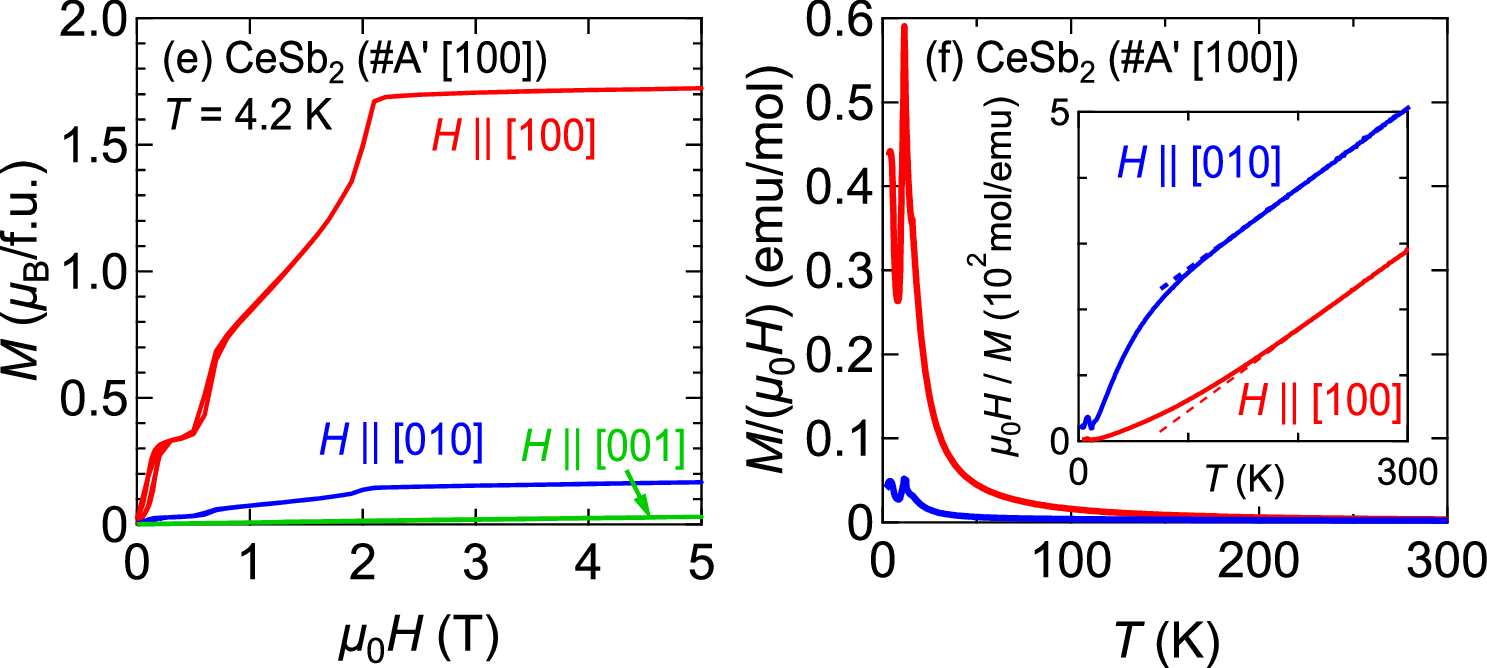}
\end{center}
\vspace{-0.5cm}
\caption{(Color online)
(a) Magnetization [$M(H)$] curves of the as-cast CeSb$_2$(\#A') at 4.2~K for $H~||$~``[100]''. (b) The $M(H)$ curves of the sample \#A' [100] for $H~||~[010]$. 
The numbers in (a) and (b) correspond to the order of measurements.
(c) The $M(H)$ curves of the sample \#A' [010] for $H~||~[100]$, measured by increasing the maximum applied fields ($\mu_0H_{\rm max}$) step by step.
The solid (dashed) lines indicate the first (second) measurement up to $\mu_0H_{\rm max}$.
The differential susceptibility $dM/d(\mu_0H)$ curves are also shown in (a)-(c).
(d) The $\mu_0H_{\rm max}$ dependence of $M_{\rm 1st}$, $M_{\rm 2nd}$, and $\Delta M$, where $M_{\rm 1st}$ ($M_{\rm 2nd}$) and $\Delta M$ are the magnetization just above the MMT for the first (second) measurement and $M_{\rm 2nd}-M_{\rm 1st}$.
The broken lines are guides.
(e) The $M(H)$ curves of the sample \#A' [100] at $T$~=~4.2~K and (f) temperature dependences of magnetic susceptibility of the sample \#A'[100] measured under 0.1~T.
The inset of (b) shows the inverse susceptibility, overlayed with the Curie-Weiss fitting results.
}
\vspace{-1.0cm}
\label{MH_a_b}
\end{figure}

Figure \ref{MH_a_b}(a) shows the high-field magnetization results [$M(H)$ curves] for the as-cast sample \#A' for $H~||$~``[100]'' at 4.2~K.
The numbers labeled on each curve indicate the order of the measurements, which we will refer to as the shot hereafter.
After showing the MMTs below 2~T, magnetization gradually increases, followed by an abrupt jump around 34~T (shot 3).
The field down-sweep magnetization shows a hysteresis loop without MMTs and becomes zero at 0~T.
The following measurements (shots 4 and 5) reveal that the magnetization value has been enhanced without any anomaly beyond the low-field MMTs.
The plausible saturated magnetization value for $H~||~[100]$ is 1.8 $\mu_{\rm B}$/f.u.
It should be noted that the differential susceptibility $dM/d(\mu_0H)$ shows a sharp and large peak surrounded by several small peaks [the inset of \ref{MH_a_b}(a)]. 
The moderate fields change the magnetic property irreversibly, similar to $R$Cu$_2$ \cite{Hashimoto_1994,Sugiyama_1997,Yoshida_1998}.
The field direction was added to the sample label to distinguish samples exposed to high magnetic fields that induce irreversible magnetization.
For example, \#A' [100] denotes the sample \#A' that was exposed to fields applied along the  ``[100]'' direction.

We subsequently continue the magnetization measurements for $H~||$~[010] with the same sample \#A' [100] by rotating the sample 90{$^\circ$} at room temperature.
Figure~\ref{MH_a_b}(b) shows the $M(H)$ curves of the sample \#A' [100] for $H~||~[010]$ at 4.2~K.
Surprisingly, the magnetization is significantly reduced from $\sim$1.5 $\mu_{\rm B}$/f.u. of the as-cast \#A' for $H~||$~``[010]'' \cite{suppl} to $\sim$0.2~$\mu_{\rm B}$/f.u. (shots 6 and 7). 
Even more surprisingly, the $M(H)$ curve (shot 8) shows similar anomalous behavior to the previous $H~||$~``[100]'' measurements for the as-cast \#A': magnetization shows a steep increase with hysteresis loop [Fig.~\ref{MH_a_b}(a)].
However, the increase in magnetization is not as abrupt as observed in the as-cast sample.
The $dM/d(\mu_0H)$ shows many peaks as if many energy barriers exist to overcome. 
Such behaviors are reminiscent of the $M(H)$ curves of the irreversible magnetization range of ferromagnets and the magnetic-field-induced martensitic transformation in metamagnetic shape memory alloys \cite{Kihara_2020}.

We again changed the field direction to [100] by remounting the sample \#A' [010] at room temperature and measured magnetization at 4.2~K [Fig.~\ref{MH_a_b}(c)].
The magnetization is again strongly suppressed to $\sim$$0.2~\mu_{\rm B}$/f.u. at 2~T: $\sim$$0.4~\mu_{\rm B}$/f.u. (the as-cast \#A') for $H~||$~``[100]'' and $\sim$$1.7~\mu_{\rm B}$/f.u. (\#A' [100]) for $H~||~[100]$ [shots 4 and 5 in Fig.~\ref{MH_a_b}(a)].
Interestingly, this magnetization value of \#A' [010] for $H~||~[100]$ just above the low-field MMT fields coincides with that of \#A' [100] for $H~||~[010]$ [shots 6 and 7 in Fig.~\ref{MH_a_b}(b)].  
Based on these findings, the magnetization easy axis of CeSb$_2$ is determined by the applied field direction. 
Specifically, the applied field direction defines the magnetization easy axis, while the magnetization hard axis is perpendicular to it.
Our magnetostriction measurements revealed that the field direction becomes the crystallographic [100] axis, which will be reported separately.
Moreover, this memory effect, observed in the ordered state, persists even at room temperature.
Note that the magnetization for $H~||~$[001] is always the lowest at least up to 50~T.
Thus, the magnetization easy-axis switch occurs within the $ab$ plane, including the Ce-pantograph networks.

To further elucidate the magnetic memory effect of CeSb$_2$, magnetization was measured by incrementally increasing the maximum applied fields, $\mu_0H_{\rm max}$, along~[100] [Fig.~\ref{MH_a_b}(c)]. 
During these measurements, the sample temperature was maintained at 4.2~K, well below $T_{\rm N}$ = 15.5~K.
The magnetization measurements were repeated twice for each $\mu_0H_{\rm max}$ value.
In the first measurement, the magnetization exhibits a steep increase beyond 25~T, accompanied by hysteresis. 
In the second measurement cycle, the magnetization approaches the maximum value attained in the first measurement without showing anomalies above 2~T.
A significant increase in magnetization was observed upon exceeding the $\mu_0H_{\rm max}$ of the previous lower-field measurements.
This effect was more pronounced in the $dM/d(\mu_0H)$.
These results suggest a unique magnetic memory effect in CeSb$_2$, where the magnetization is retained until the applied field exceeds the previously experienced field.
Finally, the magnetization reaches 1.7~$\mu_{\rm B}$/f.u for $\mu_0H_{\rm max}\approx$ 50~T.
Thus, we conclude that the magnetization easy-axis switch can be repeatedly induced by changing the applied field direction along either [100] or [010]. 
Figure \ref{MH_a_b}(d) presents the magnetization difference $\Delta M$ between the first and second measurements as a function of $\mu_0H_{\rm max}$.
A significant increase in $\Delta M$ is oberved around $\mu_0H_{\rm max}$ = 32~T.  
Above $\mu_0H_{\rm max}\approx$ 50~T, $\Delta M$ approaches zero, indicating the completion of this conversion around 50~T.

Since the memory effect persists even under ambient conditions, we can investigate the intrinsic magnetic properties of CeSb$_2$ using field-detwinned samples.
Figure \ref{MH_a_b}(e) shows the $M(H)$ curves of \#A' [100] for $H~||~$[100], [010], and [001] at 4.2~K.
Below 2~T, three-step MMTs were observed for both in-plane directions.  
For magntization easy-axis $(H~||~[100])$, the magnetization just above 2~T is 1.7~$\mu_{\rm B}$/f.u., while 0.2~$\mu_{\rm B}$/f.u. for the hard axis ($H~||~[010]$).
This difference, i.e., the magnetic anisotropy, is a key ingredient for the magnetization easy-axis conversion.
Temperature dependence of the {\it intrinsic} magnetic susceptibility of CeSb$_2$ is shown in Fig.~\ref{MH_a_b}(f).
The curves follow the Curie-Weiss law above 150~K.
The effective moment and Weiss constant for the [100] ([010]) axis are 2.60 (2.55)~$\mu_{\rm B}$ and 56 (-113)~K, respectively.
The moments for both directions are close to the Ce$^{3+}$-ion value. 
It is crucial to determine whether the crystalline electric field effect, which induces magnetic anisotropy, can account for the magnetic properties of CeSb$_2$.

These phenomena, exhibiting temperature-stable memory effects, involve both magnetic and structural changes.
We observed domain formation in the as-cast samples and field-induced domain rearrangement through the polarized light microscope images.
Figures~\ref{H_domain} (d) and (g) show the polarizing light images of the as-cast CeSb$_2$ \#C for the $c$ plane under ambient conditions.
The region with different colors, mainly bright (yellow) and dark (blue), corresponds to the different domains.
The domain pattern varies significantly among the samples and even between the surface and back sides of the same sample [compare Fig.~\ref{H_domain} (d) and (g)].
This may arise from the weak interplane connection expected from the easily cleavable $ab$ plane.  
The possible domain configurations are schematically drawn in Fig.~\ref{H_domain} (c).
The neighboring domains share diagonals of the $ac$ and $bc$ planes, referred to as the 90$^{\circ}$ domain.
The domain walls are along nearly 45$^{\circ}$ from the principal $a$ and $b$ axes due to the similarity of their lattice constants. 
The different domain distributions cause variation in the magnetization of the as-cast CeSb$_2$ samples \cite{suppl}.
The domain size is comparable to or larger than the x-ray beam diameter used for the sample orientation, meaning the identified axis is part of the sample.
Thus, the crystal axes of the as-cast bulk samples cannot be uniquely determined due to the admixture of the domains.

To investigate the impact of an external field on domain configurations, we compared the polarizing light microscope images of the as-cast CeSb$_2$ (\#C), \#C [100], and \#C [010].  
First, we confirm the reproducibility of the axis conversion by the high-field magnetization measurements.
Figure \ref{H_domain}(a) shows the $M(H)$ curves of the as-cast CeSb$_2$ (\#C) for $H~||~$``[100]'' at 4.2~K.
Magnetization exhibits an upturn above 28~T, followed by stairway-like increases in the field-up-sweep measurements (shot 2).
During the demagnetization process, magnetization monotonically decreases, similar to sample \#A' [Fig.~\ref{MH_a_b}(a)].
The magnetization value of $\sim$1.4~$\mu_{\rm B}$/f.u. just above 2~T is slightly smaller than that of the sample \#A' [100].
The polarizing light microscope images of the CeSb$_2$ \#C [100] (after shot 3) provide evidence of domain rearrangement, as shown in Fig.~\ref{H_domain} (e) and (h) [compare with Fig.~\ref{H_domain} (d) and (g)]: the light-colored region shrinks.

Next, we remount the sample \#C [100] for the $H~||~[010]$ measurements at room temperature. 
The $M(H)$ curves of sample \#C [100] for $H~||~[010]$ at 4.2~K are shown in Fig.~\ref{H_domain}(b).
These results are consistent with those obtained for sample \#A' [100] [Fig.~\ref{MH_a_b}(b)].
The smaller magnetization observed in the as-cast sample \#C for $H~||~[100]$ is also reproduced.
The domain images of sample \#C [010] (after shot 6) are shown in Figs.~\ref{H_domain}(f) and (i).
The anticipated full domain rearrangement, i.e., complete detwinning, has not been achieved in sample \#C. 
If detwinning were complete, the entire area in the images would be uniformly yellow.
This result is consistent with the reduced magnetization value.
The cascade of magnetization steps observed in the as-cast sample indicates that multiple barriers prevent the rearrangement of domains [see Fig. \ref{H_domain}(a)].
Comparing all the magnetization measurements (not shown), the axis conversion for sample \#A' (Fig.~\ref{MH_a_b}) is the most ideal.
From the magnetization curves shown in Figs.~\ref{H_domain}(a) and (b), it is inferred that approximately 80\% of the volume of sample \#C is aligned at 52~T.
 
We discuss possible scenarios explaining the axis-switch based on the unique Ce-pantograph network within the $ab$ plane (Fig. \ref{structure}).
The switch occurs for the fields along the in-plane principal axes, which coincide with the diagonals of the pantograph.
Due to the slightly larger $a$-axis lattice constant compared to the $b$-axis lattice constant, the pantograph angle along the $a$ axis is acute [Fig. \ref{structure}(c)].
The Ce-pantograph layers sandwich an Sb$_2$ layer. 
The distance between Sb atoms is comparable to the Sb$\text{--}$Sb bond length, and thus Sb$_2$ is regarded as a dimer tilting from the $c$ to $b$ axis \cite{Singha_2024}.
The switch between the $a$ and $b$ axes while maintaining the Ce$\text{--}$Ce distance is likely to occur by changing the angles of the Ce-pantograph.
The Sb$_2$ dimers between the pantograph layers may behave as molecules in a solid.
The stretching/compressing of the pantograph accompanies the tilting of the Sb$_2$-dimers. 
Thus, this structural change, i.e., the domain rearrangement, is not diffusive like a typical structural transition, but rather a diffusionless transformation, similar to the martensite transformation.

We observed the domain formation under ambient conditions using the polarizing light microscope.
Generally, the structural transitions that lower the crystal symmetry lead to twinning in the sample.  
Thus, the tetragonal phase may exist at high temperatures and ambient pressure. 
The as-cast samples, obtained under ambient conditions, have undergone a structural transition, according to the temperature-pressure phase diagram of CeSb$_2$\cite{Kagayama_2000, Squire_2023, Weinberger_2022, Hodgson_2023}.
However, the high-pressure phase, which connects to the high-temperature phase at ambient pressure, is proposed to exhibit orthorhombic symmetry ($Cmcm$) \cite{Squire_2023}.
A similar pressure-induced structural transition in the sister compound SmSb$_2$ was reported, from orthorhombic ($Cmce$) to tetragonal ($P4/nmm$) \cite{Li_2024}.
It is crucial to elucidate the high-temperature crystal structure of CeSb$_2$, as it may influence the high-pressure phase in CeSb$_2$, including magnetism and superconductivity \cite{Squire_2023}. 
In addition, the previous reports have been conducted using the as-cast ``twinned'' samples \cite{Budko_1998, Zhang_2017, Liu_2020, Trainer_2021,Kagayama_2000,Squire_2023}. 
They should be revisited by using detwinned samples.

The axis-conversion phenomena were originally observed in $R$Cu$_2$ \cite{Hashimoto_1994}.
From a structural point of view, slight distortion from higher symmetry may be an essential ingredient for the axis conversion.
Since the crystal structure of $R$Cu$_2$ is derived from an AlB$_2$-type
hexagonal structure, three variants, turning the propagation almost 120$^{\circ}$ from each other, are observed \cite{Raasch_2006}.
A change in the volume fraction of the twin variants with the applied fields has been observed, providing an alternative scenario to quadrupole moment flipping \cite{Settai_1995, Yoshida_1998}.
UPtGe also crystallizes in the pseudo-hexagonal structure, slightly distorted AlB$_2$-type. 
However, its first-order MMTs are reversible \cite{Miyake_2018}.
The crystal structure of CeSb$_2$ is distorted tetragonal, leading to the unique Ce-pantograph network.
In addition to the pseudo-tetragonal symmetry, the Ce-pantograph networks, which have significant magnetic anisotropy, play a crucial role in the axis conversion of CeSb$_2$.
To elucidate the role of magnetic anisotropy in the axis conversion, studies on other $R$Sb$_2$ series are important and are in progress.

In summary, we have discovered a magnetic-field-induced magnetization easy-axis switch, accompanied by domain rearrangement, in CeSb$_2$ above 30~T.
This phenomenon represents a magnetic shape memory effect that persists below room temperature. 
The direction along (perpendicular to) the magnetic field that induces the conversion becomes the magnetization easy (hard) axis.
This conversion can be repeatedly induced by changing the field direction along the in-plane principal axes.
The unique in-plane Ce-pantograph networks with significant magnetic anisotropy are key ingredients for the magnetization easy-axis switching.

\begin{acknowledgment}
We thank T. Aoyama, M. Gen, H. Sakai, H. Kotegawa, K. Kudo, H. Harima, Y. $\bar{\rm O}$nuki, and K. Kindo for fruitful discussions.
This work was carried out under the Visiting Researcher's Program of ISSP, the University of Tokyo.
This work was supported by KAKENHI (JP19K21840, JP20K03827, JP22K03516, JP22H04933, JP24H01641, JP23H04862, JP23H04870, and JP24H01601).
\end{acknowledgment}

\newpage

\centerline{\Large{\bf Supplemental material}}

%
\renewcommand{\thefigure}{S\arabic{figure}}


\section{Sample dependence of magnetization}

\begin{figure}[h]
\begin{center}
\includegraphics[width=1\hsize]{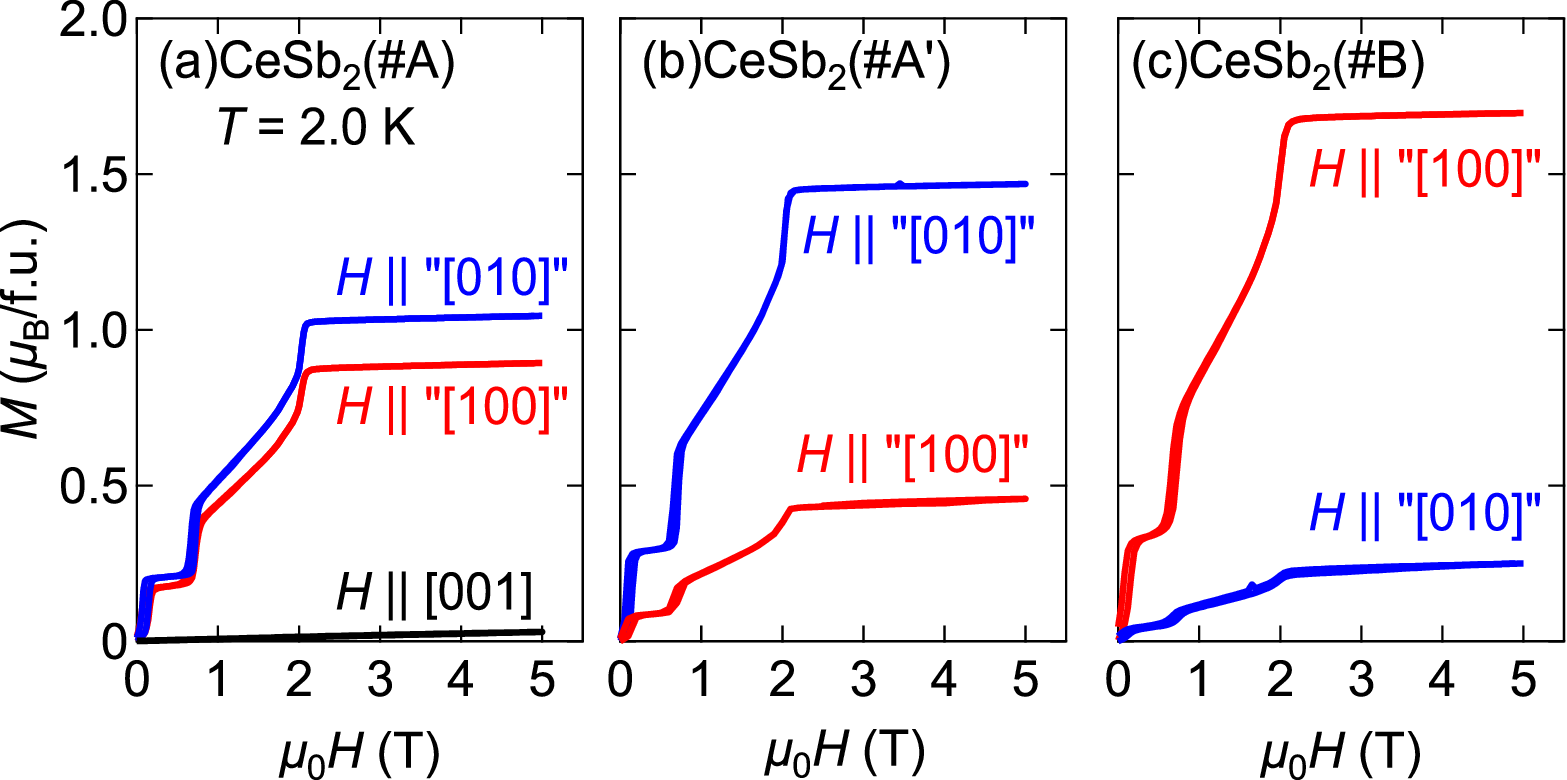}
\end{center}
\caption{(Color online)
Sample dependence of magnetization of the as-grown CeSb$_2$ at 2.0~K.
(a), (b) and (c) correspond to the results obtained for the as-cast samples \#A, \#A', and \#B, respectively.
}
\label{MPMS}
\end{figure}

Magnetization of as-cast CeSb$_2$ depends on samples (Fig. \ref{MPMS}).
The metamagnetic transition (MMT) fields along the [100] and [010] axes coincide, reflecting the pseudo-tetragonal symmetry.
Furthermore, the MMT fields remain consistent across the studied samples.
However, the absolute magnetization values differ between samples, even for those cut from the same piece (samples \#A and \#A').
This sample dependence is attributed to twinning in the as-cast samples.

\end{document}